# Artificial Rydberg Atom

Yong S. Joe
Center for Computational Nanoscience
Department of Physics and Astronomy
Ball State University
Muncie, IN 47306

Vanik E. Mkrtchian
Institute for Physical Research
Armenian Academy of Sciences
Ashtarak-2, 378410
Republic of Armenia

Sun H. Lee
Center for Computational Nanoscience
Department of Physics and Astronomy
Ball State University
Muncie, IN 47306

## Abstract

We analyze bound states of an electron in the field of a positively charged nanoshell. We find that the binding and excitation energies of the system decrease when the radius of the nanoshell increases. We also show that the ground and the first excited states of this system have remarkably the same properties of the highly excited Rydberg states of a hydrogen-like atom i.e. a high sensitivity to the external perturbations and long radiative lifetimes.

# I. Introduction

The study of nanophysics has stimulated a new class of problem in quantum mechanics and developed new numerical methods for finding solutions of many-body problems [1]. Especially, enormous efforts have been focused on the investigations of nanosystems where electronic confinement leads to the quantization of electron energy. Modern nanoscale technologies have made it possible to create an artificial quantum confinement with a few electrons in different geometries.

Recently, there has been a discussion about the possibility of creating a new class of spherical artificial atoms, using charged dielectric nanospheres which exhibit charge localization on the exterior of the nanosphere [2]. In addition, it was reported by Han *et al* [3] that the charged gold nanospheres were coated on the outer surface of the microshells. They have shown that the variations of absorption spectrum in the gold nanosphere stem from the enhanced particle interactions and from the field enhancements in the space among the nanospheres. The crucial role of charged nanospheres in photoactive nanocompounds has been emphasized in Ref. [4].

In the meantime, Rydberg atoms i.e. hydrogen-like atoms in which the external electron is excited to a state of a large principal quantum number have been studied intensively in the area of laser spectroscopy [5]. The average separation between the excited electron and core ion is too large and hence, the external electron is loosely bound with the nucleus. For instance, it is reported that the excited electron in a quantum state with a principal quantum number $n = 27$ has an average distance of $37nm$ from the nucleus [6]. Due to their large size, Rydberg atoms exhibit large diamagnetic energy shifts. As a consequence, Rydberg atoms are highly sensitive to the external perturbations which make them very important in different applications [5].

However, one of the restrictions on the Rydberg atoms in the application point of view is a necessity of pumping systems (lasers). In this article we propose a new nanosystem, called *artificial Rydberg atoms* (ARA), which has surprisingly the same properties of the Rydberg atom, but does not need a pumping system. This ARA can simply be achieved by considering a system of an electron in the field of a positively charged nanoshell. Hence, we study a simple model of the charged nanoshell potential and examine a comprehensive description of the quantum mechanics in this system. These investigations allow us to develop an idea of how the spectrum and distribution of electron change with a radius of the nanoshell and to find a deep analogy with Rydberg atoms. Besides, they provide the first step towards many-electron theory of a metamaterial consisting of ARA-s.

The paper is organized as follows. In Sec. II, we describe a model of nanoshell potential. Section III contains general expressions for the solution of eigenvalue and eigenfunction problems in the field of a potential introduced in Sec. II. It is also shown that the results are in good agreement with the well-known solution of the problem for an electron in the spherical symmetric rectangular well with a finite depth. In Sec. IV, we analyze electron energy levels and eigenfunctions in a charged nanoshell potential, and show that we come to the results for a hydrogen atom in the limit when the radius of the nanoshell approaches to zero. We present numerical results in Sec. V on the electron energy levels, excitation energies, electron distribution functions, and matrix elements for different radii of the nanoshell and different electron orbital quantum number. Finally, we summarize our concluding remark in Sec. VI and calculate the spectrum of ARA in Appendix A using the WKB approximation.

## II. Model of nanoshell

We study a simple model of nanoshell by ignoring thickness and other possible structures of the nanoshell, because the inclusion of a finite thickness of the shell does not bring to the new qualitative consequences. We consider an electron in the field of a spherically symmetric potential:

$$V(r) = \begin{cases} -V_0, & r \leq R \\ -\frac{\alpha}{r}, & r \geq R \end{cases}, \quad (1)$$

where $V_0$ and $\alpha$ are positive constant numbers:

$$V_0, \alpha \geq 0. \quad (2)$$

Here, Eq. (1) is coincident with the potential of a spherically symmetric rectangular well with a depth $V_0$ when $\alpha = 0$. For a spherical shell having the magnitude of a positive charge $Z|e|$, we have a constant $\alpha$ as

$$\alpha = Ze^2, \qquad (3)$$

and a constant $V_0$ as

$$V_0 = \frac{Ze^2}{R} \qquad (4)$$

with a condition that the electrostatic potential $V(r)$ is continuous at $r=R$. For the sake of simplicity, we have ignored an influence of the medium in Eq. (1). In other words, an electron in the field of a spherically symmetric potential is assumed to be in vacuum.

### III. Bound states

In this section, we focus on determining the bound energy levels and eigenfunctions of the electron in the field of a potential given in Eq. (1), that is E<0 and

$$|E| \leq V_0. \qquad (5)$$

Defining quantities

$$\chi = \sqrt{\frac{2m}{\hbar^2}|E|}, \qquad (6.a)$$

$$k = \sqrt{\frac{2m}{\hbar^2}(V_0 - |E|)}, \qquad (6.b)$$

$$\lambda = \frac{m\alpha}{\chi \hbar^2}, \qquad (6.c)$$

the radial Schrödinger equation [7] in the field of a potential [Eq. (1)] can be written as

$$\Phi^{<"}(r) + \frac{2}{r}\Phi^{<'}(r) + \left[k^2 - \frac{l(l+1)}{r^2}\right]\Phi^{<}(r) = 0 \qquad (7.a)$$

for $r < R$, and

$$\Phi^{>"}(r) + \frac{2}{r}\Phi^{>'}(r) - \left[\chi^2 - 2\frac{\chi\lambda}{r} + \frac{l(l+1)}{r^2}\right]\Phi^{>}(r) = 0 \qquad (7.b)$$

for $r > R$. Since $\Phi^{<}(r)$ has to be finite for $r = 0$, a solution of the Eq. (7.a) can be obtained as

$$\Phi^{<}(r) = A_l j_l(kr), \qquad (8.a)$$

where $A_l$ is a constant and $j_l(kr)$ is the first kind of a spherical Bessel function [8]. In the meantime, one can obtain a solution of Eq. (7.b) as

$$\Phi^>(r) = B_l(\chi r)^l e^{-\chi r} U(l+1-\lambda, 2l+2, 2\chi r), \tag{8.b}$$

where $B_l$ is a constant and $U(a,b,x)$ is the second kind of a confluent function [8]. Both wave functions and the first derivative of the wave functions have to be continuous at the boundary $r = R$:

$$\Phi^<(R) = \Phi^>(R), \quad \Phi^{<'}(R) = \Phi^{>'}(R). \tag{9}$$

Then, from Eq. (8) and Eq. (9), we get an expression for coefficients

$$B_l = \frac{(\chi R)^{-l} e^{\chi R} j_l(kR)}{U(l+1-\lambda, 2l+2, 2\chi R)} A_l \tag{10}$$

and a transcendental equation for the bound state energy levels:

$$k \frac{j_l'(kR)}{j_l(kR)} + \chi - \frac{l}{R} - 2\chi \frac{U'(l+1-\lambda, 2l+2, 2\chi R)}{U(l+1-\lambda, 2l+2, 2\chi R)} = 0. \tag{11}$$

Here, we note that differentiation of the special functions $j_l$ and $U$ is taken by the whole arguments.

Therefore, the general wave function can be obtained from Eq. (8) and Eq. (10) as

$$\Phi(r) = C \left[ \theta(R-r) \frac{j_l(kr)}{j_l(kR)} + \theta(r-R) \left(\frac{r}{R}\right)^l e^{-\chi(r-R)} \frac{U(l+1-\lambda, 2l+2, 2\chi r)}{U(l+1-\lambda, 2l+2, 2\chi R)} \right], \tag{12}$$

where $C$ is defined from a normalization condition:

$$\int_0^{+\infty} dr\, r^2 \Phi^2(r) = 1. \tag{13}$$

By taking $\alpha = 0$ in the potential of Eq. (1), we come to the problem of an electron in the spherically symmetric rectangular well with a finite depth. In this case, $\lambda = 0$ [see (6.c)] and then a confluent function $U(l+1, 2l+2, 2\kappa R)$ may be expressed by the first kind of a Hankel function [8]:

$$U(l+1, 2l+2, 2\chi R) = \frac{\sqrt{\pi}}{2}(-1)^{l+1}(2i\chi R)^{-l-1/2} e^{\chi R} H^{(1)}_{l+1/2}(i\chi R).$$

Hence, Eq. (11) and Eq. (12) becomes:

$$k \frac{j_l'(kR)}{j_l(kR)} - i\chi \frac{h_l^{(1)\prime}(i\chi R)}{h_l^{(1)}(i\chi R)} = 0,$$

$$\Phi(r) = C \left[ \theta(R-r) \frac{j_l(kr)}{j_l(kR)} + \theta(r-R) \left(\frac{r}{R}\right)^{-l} \frac{h_l^{(1)}(i\chi r)}{h_l^{(1)}(i\chi R)} \right],$$

where $h_l^{(1)}(x)$ is the first kind of a spherical Hankel function [8]. These expressions for the energy spectrum and the wave functions of a particle in the field of a spherically symmetric rectangular well potential are known in the quantum theory of nucleus [9].

### IV. Electron energy levels and eigenfunctions

Let us analyze, in detail, the case when the potential of Eq. (1) is coincident with a scalar potential of the charged spherical shell [Eq. (3) and Eq. (4)] with $Z = 1$. We use Coulomb units by taking

$$|E| = \frac{1}{2} \xi^2 E_a \quad \text{and} \quad R = \eta a_0, \tag{14}$$

where $E_a = me^4/\hbar^2 = 2Ry = 27.21 eV$ and $a_0 = \hbar^2/me^2 = 0.053 nm$ are Coulomb units of an energy (Rydberg) and a distance (Bohr radius), respectively. Because $\eta$ is changed in a region $0 \leq \eta < +\infty$ and the energy is satisfied from an inequality relation of Eq. (5), a condition for $\xi$ is determined by

$$0 \leq \xi \leq \sqrt{2/\eta}. \tag{15}$$

Inserting Eq. (14) into the dispersion relation of Eq. (11), we find:

$$\beta \frac{j_l'(\beta\xi\eta)}{j_l(\beta\xi\eta)} + 1 - \frac{l}{\xi\eta} - 2 \frac{U'(l+1-\xi^{-1}, 2l+2, 2\xi\eta)}{U(l+1-\xi^{-1}, 2l+2, 2\xi\eta)} = 0, \tag{16}$$

where $\beta = \sqrt{2(\eta\xi^2)^{-1} - 1}$. Using differentiation rules for the Bessel and confluent functions such as $j_l'(x) = \frac{l}{x} j_l(x) - j_{l+1}(x)$ and $U'(a,b,x) = U(a,b,x) - U(a,b+1,x)$, Eq. (16) can be expressed by

$$U(l+1-\xi^{-1}, 2l+3, 2\xi\eta) - \Lambda_l(\xi,\eta) U(l+1-\xi^{-1}, 2l+2, 2\xi\eta) = 0, \tag{17}$$

where

$$\Lambda_l(\xi,\eta) \equiv \frac{1}{2}\left[1+\beta\frac{j_{l+1}(\beta\xi\eta)}{j_l(\beta\xi\eta)}\right]. \tag{18}$$

It is interesting to note that in this system, there is no accidental degeneracy of energetic levels which is different from that of a pure Coulomb potential case (a hydrogen atom). The parameter $\xi$ in solutions of Eq. (17) is a function of the nanoshell radius $\eta$. In addition, it also depends on the orbital quantum number $l$ and the radial quantum number $n_r = 0, 1, 2, ...$:

$$\xi = \xi_{l,n_r}(\eta). \tag{19}$$

Inserting Eq. (19) and Eq. (14) into Eq. (12), we find an eigenfunction

$$\Phi_{l,n_r}(\rho) = a_0^{-3/2} C_{l,n_r} \begin{cases} \dfrac{j_l(\lambda_{l,n_r}\rho)}{j_l(\lambda_{l,n_r}\eta)} & \text{for } \rho < \eta \\ \dfrac{\rho^l e^{-\xi_{l,n_r}\rho} U(l+1-\xi_{l,n_r}^{-1}, 2l+2, 2\xi_{l,n_r}\rho)}{\eta^l e^{-\xi_{l,n_r}\eta} U(l+1-\xi_{l,n_r}^{-1}, 2l+2, 2\xi_{l,n_r}\eta)} & \text{for } \rho > \eta \end{cases} \tag{20}$$

with a corresponding eigenvalue

$$E_{l,n_r} = -\xi_{l,n_r}^2 Ry. \tag{21}$$

Here, we have introduced new parameters $\lambda_{l,n_r} = \sqrt{2/\eta - \xi_{l,n_r}^2}$ and $\rho \equiv r/a_0$ in Eq. (20). The coefficient $C_{l,n_r}$ in Eq. (20) can be obtained from the normalization condition of Eq. (13):

$$C_{l,n_r}^{-2} = \int_0^\eta d\rho \rho^2 \left[\frac{j_l(\lambda_{l,n_r}\rho)}{j_l(\lambda_{l,n_r}\eta)}\right]^2 + \int_\eta^{+\infty} d\rho \rho^2 \left[\frac{\rho^l e^{-\xi_{l,n_r}\rho} U(l+1-\xi_{l,n_r}^{-1}, 2l+2, 2\xi_{l,n_r}\rho)}{\eta^l e^{-\xi_{l,n_r}\eta} U(l+1-\xi_{l,n_r}^{-1}, 2l+2, 2\xi_{l,n_r}\eta)}\right]^2. \tag{22}$$

Now, we would like to confirm that if we consider a limiting case of the above results, then they are in good agreement with the solution of a hydrogen atom problem. In other words, we calculate a dispersion relation of Eq. (17) and wave functions of Eq. (20) in the limit $\eta \to 0$, which is a pure Coulomb potential in Eq. (1) with $R = 0$. Using an expression $\dfrac{j_{l+1}(x)}{j_l(x)} = 2x \sum_{n=1}^{\infty} \dfrac{1}{\gamma_{l+1/2,n}^2 - x^2}$ [10] where $\gamma_{l,n}$ ($n = 0, 1, 2, ...$) are positive zeros of the Bessel function $j_l(x)$, $\Lambda_l(\xi,\eta)$ in Eq. (18) can be written as

$$\Lambda_l(\xi,\eta) = \frac{1}{2} + \beta^2 \xi\eta \sum_{n=1}^{\infty} \frac{1}{\gamma_{l+1/2,n}^2 - (\beta\xi\eta)^2}.$$

Hence, the magnitude of $\Lambda_l(\xi,\eta)$ at $\eta = 0$ in Eq. (18) can be obtained as

$$\Lambda_l(\xi,0) = \frac{1}{2} + \frac{2}{\xi}\sum_{n=1}^{\infty}\frac{1}{\gamma_{l+1/2,n}^2} = \frac{1}{2} + \frac{1}{\xi(2l+3)}, \quad (23)$$

where the sum over zeros of the Bessel function is expressed by $\sum_{n=1}^{\infty}\frac{1}{\gamma_{l+1/2,n}^2} = \frac{1}{2(2l+3)}$ [10].

On the other hand, the confluent function $U(a,b,x)$ for $x \to +0$ and $b > 2$ is given by [8]

$$U(a,b,x) \to \frac{\Gamma(b-1)}{\Gamma(a)}\frac{1}{x^{b-1}}. \quad (24)$$

Using Eq. (23) and Eq. (24), we find the dispersion relation of Eq. (17) that is satisfied if and only if the condition

$$\frac{1}{\Gamma(l+1-\xi^{-1})} = 0 \quad (25)$$

is provided. Taking into account the fact that $\Gamma$ function has poles at non-positive integers [8], we extract $l+1-\xi^{-1} = -n_r$ $(n_r = 0,1,2,......)$ from Eq. (25). Therefore, we find the solutions of Eq (19) at $\eta = 0$ as

$$\xi_{l,n_r}(0) = \frac{1}{n_r + l + 1}. \quad (26)$$

Inserting Eq. (26) into Eq. (21), we finally have the spectrum of a hydrogen atom:

$$E_{l,n_r} = -\frac{Ry}{(n_r+l+1)^2}. \quad (27)$$

Substituting Eq. (26) into Eq. (20) and taking the limit $\eta \to 0$, we obtain a coordinate dependent part of wave function as

$$\rho^l e^{-\xi_{l,n_r}\rho}U(-n_r, 2l+2, 2\xi_{l,n_r}\rho).$$

After we use the confluent function $U$ in terms of the Laguerre polynomials, $U(-n,\alpha+1,x) = (-1)^n n! L_n^{(\alpha)}(x)$ for $n = 0,1,2,...$, we get the radial wave function of a hydrogen atom [7]. We notice here that if we change $\xi \to Z\xi$ and $\eta \to Z^{-1}\eta$, then we can generalize our results in the case of $Z \neq 1$.

## V. Analogy with Rydberg states

In this section, we analyze electronic energy levels, excitation energies and electron distribution functions in the ARA for different radii of the nanoshell. The first four levels, $\xi_{l,n_r}(\eta)$, for $l = 0, 1, 2$ and $\eta = 50$ $(R = 2.65nm)$, $100$ $(R = 5.3nm)$, $150 (R = 7.95nm)$, and $200 (R = 10.6nm)$ are listed in Table 1 by solving Eq. (17). The energy of the electron in ARA is then calculated from Eq. (21). As in case of Rydberg atoms [5], the binding energies of ARA are an order of $10^{-2} Ry$. As we can see from Table 1, $\xi$ -s are monotonic decreasing functions of $\eta$. For example, $\xi_{0,0}(0) = 1$ for a hydrogen atom $(\eta = 0)$ and $\xi_{0,0}(\eta)$ at $\eta = 50, 100, 150, 200$ for a nanoshell is equal to 0.19429, 0.13917, 0.11418, and 0.09913, respectively. The monotonic decreasing behavior of $\xi_{l,n_r}(\eta)$ for the ground and the first excited levels of ARA is analytically shown in Appendix A using the WKB approximation. We also note from Table 1 that the energy levels in ARA have the following arrangement for a fixed radius of the nanoshell as

$$E_{0,n_r} < E_{1,n_r} < E_{2,n_r} < .... < E_{0,n_r+1} < E_{1,n_r+1} < E_{2,n_r+1} < .....$$

This clearly indicates that there is no accidental degeneracy in ARA, which is not the case for a pure Coulomb potential (a point nucleus).

| $\eta=50$ | $l=0$ | $l=1$ | $l=2$ | $\eta=100$ | $l=0$ | $l=1$ | $l=2$ |
|---|---|---|---|---|---|---|---|
| $n_r=0$ | 0.19429 | 0.18832 | 0.18078 | $n_r=0$ | 0.13917 | 0.13682 | 0.13385 |
| $n_r=1$ | 0.17771 | 0.16730 | 0.15675 | $n_r=1$ | 0.13252 | 0.12810 | 0.12322 |
| $n_r=2$ | 0.15594 | 0.14673 | 0.13783 | $n_r=2$ | 0.12230 | 0.11701 | 0.11206 |
| $n_r=3$ | 0.13782 | 0.12942 | 0.12197 | $n_r=3$ | 0.11184 | 0.10708 | 0.10227 |
| $\eta=150$ | $l=0$ | $l=1$ | $l=2$ | $\eta=200$ | $l=0$ | $l=1$ | $l=2$ |
| $n_r=0$ | 0.11418 | 0.11283 | 0.11113 | $n_r=0$ | 0.09913 | 0.09823 | 0.09708 |
| $n_r=1$ | 0.11034 | 0.10776 | 0.10486 | $n_r=1$ | 0.09655 | 0.09480 | 0.09282 |
| $n_r=2$ | 0.10421 | 0.10076 | 0.09733 | $n_r=2$ | 0.09235 | 0.08992 | 0.08740 |
| $n_r=3$ | 0.09702 | 0.09371 | 0.09048 | $n_r=3$ | 0.08709 | 0.08452 | 0.08207 |

TABLE 1. The first four energy levels of ARA for $l = 0, 1, 2$ and $\eta = 50, 100, 150, 200$.

For the application point of view, the excitation energy $\Delta E(\eta)$ of the system, which is the difference between the first excited and ground energy state levels, is very important. From Table 1, we can calculate the excitation energy as follows:

$$\Delta E(50) = 22.8 \cdot 10^{-4} Ry; \quad \Delta E(100) = 6.5 \cdot 10^{-4} Ry;$$
$$\Delta E(150) = 3.1 \cdot 10^{-4} Ry; \quad \Delta E(200) \cong 1.8 \cdot 10^{-4} Ry. \tag{28}$$

This indicates that a transition frequency $\Delta \omega = \Delta E / \hbar$, which corresponds to the excitation energy, is an order of THz in the whole range of the change of the nanoshell radius 2.65nm-10.6nm. Because the radiative lifetime of the excited level is proportional to $(\Delta \omega)^3$, we may conclude that the first excited levels of ARA also have very long lifetimes analogous of the Rydberg atom.

In order to examine the properties of highly excited Rydberg states of the ARA, such as a high sensitivity to the external perturbations, we study an electron distribution function defined as

$$D_{l,n_r}(\rho) = \rho^2 \Phi_{l,n_r}^2(\rho). \tag{29}$$

Figure 1 shows the electron distribution function of a nanoshell with $R = 10.6 nm$ ($\eta = 200$) and $l = 0$ for (a) $n_r = 0$, (b) $n_r = 1$, (c) $n_r = 2$, and (d) $n_r = 3$. It is clearly seen that the electron clouds (i.e. distribution of electrons around the nucleus) in ARA ($\rho \approx 300$) is approximately a hundred times larger than that of a hydrogen atom ($\rho \approx 3$). Since the electron clouds are directly related to the dipole matrix elements as shown in the Rydberg atoms, a large distribution of electrons induces huge dipole matrix elements and makes a high sensitivity of the system to external perturbations.

Because the diamagnetic response of the system is proportional to the magnitude of the matrix element, we now estimate the magnitude of diamagnetic susceptibility of ARA by taking a matrix element of $r^2$ in the ground state. The diamagnetic susceptibility $\chi$ at temperature $T = 0$ is proportional to the area of electron clouds in the ground state $|G\rangle$ [7]:

$$\chi \sim \langle G|r^2|G\rangle. \tag{30}$$

Using Eq. (20) for wave functions, we calculate $\langle G|r^2|G\rangle$ for different radii of the nanoshell $\eta = 50,100,150$ and 200 and find out:

$$\langle G|r^2|G\rangle = \begin{cases} 1292.3a_0^2 & \text{for } \eta = 50 \\ 4527.9a_0^2 & \text{for } \eta = 100 \\ 9573.2a_0^2 & \text{for } \eta = 150 \\ 16373.5a_0^2 & \text{for } \eta = 200 \end{cases}. \tag{31}$$

This result indicates that the magnitude of the diamagnetic susceptibility of ARA increases with the radius of nanoshell and is one–ten thousand times larger than that of the hydrogen atom. In other words, a large diamagnetic response at $T = 0$ in ARA is obtained due to a large matrix element, as expected for Rydberg atoms.

## VI. Conclusion

In summary, we have studied the quantum mechanics of an electron in the field of a positively charged nanoshell. In this system, we have found that as the radius of the nanoshell increases, the size of the electron clouds increases, but the binding and excitation energies of the system decrease. We have also shown that the ground and the first excited bound states of the system have remarkably the same properties of highly excited Rydberg states of a hydrogen-like atom, such as a high sensitivity to the external perturbations and long radiative lifetimes. Finally, we have also calculated the ground and the first excited state energies using the WKB approximation. It is confirmed from both the exact results and the WKB approximation that the magnitudes of the ground and the first excited state energies are in good agreement.

## APPENDIX A: ARA spectrum in the WKB approximation

It is well-known that the WKB theory works well for a hydrogen atom and the WKB theory of a hydrogen atom gives a correct result for the energy spectrum [9]. In general, the energy of the system in the WKB approximation is given by an equation:

$$\int_{r_1}^{r_2} dr Q(r) = \pi(n_r + 1/2); \quad n_r = 0,1,2,3,... \tag{A.1}$$

where

$$Q^2(r) = \frac{2m}{\hbar^2}\left[E - V(r) - \frac{\hbar^2}{2mr^2}(l+1/2)^2\right] \qquad (A.2)$$

and $r_{1,2}$ are classical turning points with a condition:

$$Q(r_{1,2}) = 0. \qquad (A.3)$$

There are two types of the finite motion depending on the energy of the particle in the field of the symmetric potential [Eq. (1)]. If the energy $\xi^2$ of the particle in Eq. (14) is satisfied with an inequality relation

$$\xi^2 \leq \min\left[\frac{1}{L^2},\ 2/\eta - (L/\eta)^2\right] \qquad (A.4)$$

and

$$L \leq \sqrt{2\eta}, \qquad (A.5)$$

then there are two turning points: one is $\rho^<$ which is inside the nanoshell and another one is $\rho_+^>$ which is outside the nanoshell. Hence, the momentum of the particle at the point $\rho$ can be written as

$$Q(\rho) = \begin{cases} [2/\eta - \xi^2 - L^2/\rho^2]^{1/2} & \text{for } \rho^< \leq \rho \leq \eta, \\ [2/\rho - \xi^2 - L^2/\rho^2]^{1/2} & \text{for } \eta \leq \rho \leq \rho_+^>, \end{cases} \qquad (A.6)$$

where we introduce

$$L = l + 1/2. \qquad (A.7)$$

Therefore, the turning points in the problem can be obtained:

$$\rho^< = L/\sqrt{2/\eta - \xi^2},\ \rho_\pm^> = \xi^{-2}\left[1 \pm \sqrt{1-(\xi L)^2}\right]. \qquad (A.8)$$

For the energy window of particle satisfying an inequality relation

$$2/\eta - (L/\eta)^2 \leq \xi^2 \leq \min\left[1/\eta,\ 1/L^2\right] \qquad (A.9)$$

both turning points $\rho_\pm^>$ are located outside of the nanoshell and the momentum of the particle is given by

$$Q(\rho) = [2/\rho - \xi^2 - L^2/\rho^2]^{1/2} \text{ for } \rho_-^> \leq \rho \leq \rho_+^>. \qquad (A.10)$$

Substituting Eq. (A.5) and Eq. (A.7) into Eq. (A.10) and taking integration by $\rho$, we come to the result as follows. In the WKB approximation the energy levels $\xi = \xi_{l,n_r}$ of ARA shows a hydrogen-like spectrum such as

$$\xi_{l,n_r} = \frac{1}{l+n_r+1}; \quad n_r = 0,1,2,... \qquad (A.11.a)$$

in the energy window

$$2/\eta - (L/\eta)^2 \le \xi^2 \le \min[1/\eta,\ 1/L^2]. \qquad (A.11.b)$$

Furthermore, the spectrum of ARA is obtained from a transcendental equation

$$\arcsin\frac{1-L^2/\eta}{\sqrt{1-(\xi L)^2}} - \arccos\frac{L}{\sqrt{2\eta-(\xi\eta)^2}}$$

$$+ \frac{1}{\xi L}\arcsin\frac{1-\xi^2\eta}{\sqrt{1-(\xi L)^2}} - \frac{\pi}{L}\left[n_r + (L+1-\xi^{-1})/2\right] = 0 \qquad (A.12.a)$$

with a condition for an energy

$$\xi^2 \le \min\left[\frac{1}{L^2},\ 2/\eta - (L/\eta)^2\right] \qquad (A.12.b)$$

and a momentum

$$l \le \sqrt{2\eta} - 1/2. \qquad (A.12.c)$$

The solutions of Eq. (A.12.a) are shown in Table 2 for the same energetic levels as in Table 1.

| $\eta=50$ | $l=0$ | $l=1$ | $l=2$ | $\eta=100$ | $l=0$ | $l=1$ | $l=2$ |
|---|---|---|---|---|---|---|---|
| $n_r=0$ | 0.19497 | 0.18824 | 0.17992 | $n_r=0$ | 0.13960 | 0.13705 | 0.13379 |
| $n_r=1$ | 0.17693 | 0.16694 | 0.15696 | $n_r=1$ | 0.13224 | 0.12772 | 0.12296 |
| $n_r=2$ | 0.15621 | 0.14665 | 0.13759 | $n_r=2$ | 0.12221 | 0.11713 | 0.11214 |
| $n_r=3$ | 0.13772 | 0.12953 | 0.12186 | $n_r=3$ | 0.11184 | 0.10696 | 0.10228 |
| $\eta=150$ | $l=0$ | $l=1$ | $l=2$ | $\eta=200$ | $l=0$ | $l=1$ | $l=2$ |
| $n_r=0$ | 0.11447 | 0.11305 | 0.11121 | $n_r=0$ | 0.09935 | 0.09841 | 0.09719 |
| $n_r=1$ | 0.11025 | 0.10753 | 0.10461 | $n_r=1$ | 0.09653 | 0.09467 | 0.09264 |
| $n_r=2$ | 0.10403 | 0.10073 | 0.09741 | $n_r=2$ | 0.09219 | 0.08983 | 0.08740 |
| $n_r=3$ | 0.09711 | 0.09371 | 0.09040 | $n_r=3$ | 0.08713 | 0.08458 | 0.08206 |

TABLE 2. The first four energy levels of ARA for $l = 0,1,2$ and $\eta = 50,100,150,200$ in the WKB approximation.

It is worthwhile to mention that from the comparison of Tables 1 and 2, there is a small deviation (an order of $\sim 10^{-4}$) in the values of $\xi$ between the exact calculations and the WKB approximation.

With the help of Eq. (A.12.a) it is possible to find $\xi'(\eta)$ using an implicit function theorem:

$$\xi'(\eta) = -F'_\eta / F'_\xi, \qquad (A.13)$$

where $F = F(\xi, \eta)$ is coincident with the left hand side of Eq. (A.12.a). From the calculations, we get

$$\xi'(\eta) = -\frac{\xi}{2\eta}\left[1 + \frac{\xi^{-1} - \eta\xi/2}{\sqrt{2\eta - L^2 - (\xi\rho)^2}}\left(\frac{\pi}{2} + \arcsin\frac{1 - \xi^2\eta}{\sqrt{1 - (\xi L)^2}}\right)\right]^{-1}. \qquad (A.14)$$

Using an inequality relation of Eq. (15), we can conclude that

$$\xi'(\eta) < 0, \qquad (A.15)$$

and therefore $\xi_{l,n_r}$ is a monotonic decreasing function of $\eta$ in the intervals (A.12.b) and (A.12.c).

## Figure captions

Fig. 1. The electron distribution function of ARA with $R = 10.6 nm \left(\eta = 200\right)$ and $l = 0$ for (a) $n_r = 0$, (b) $n_r = 1$, (c) $n_r = 2$, and (d) $n_r = 3$.

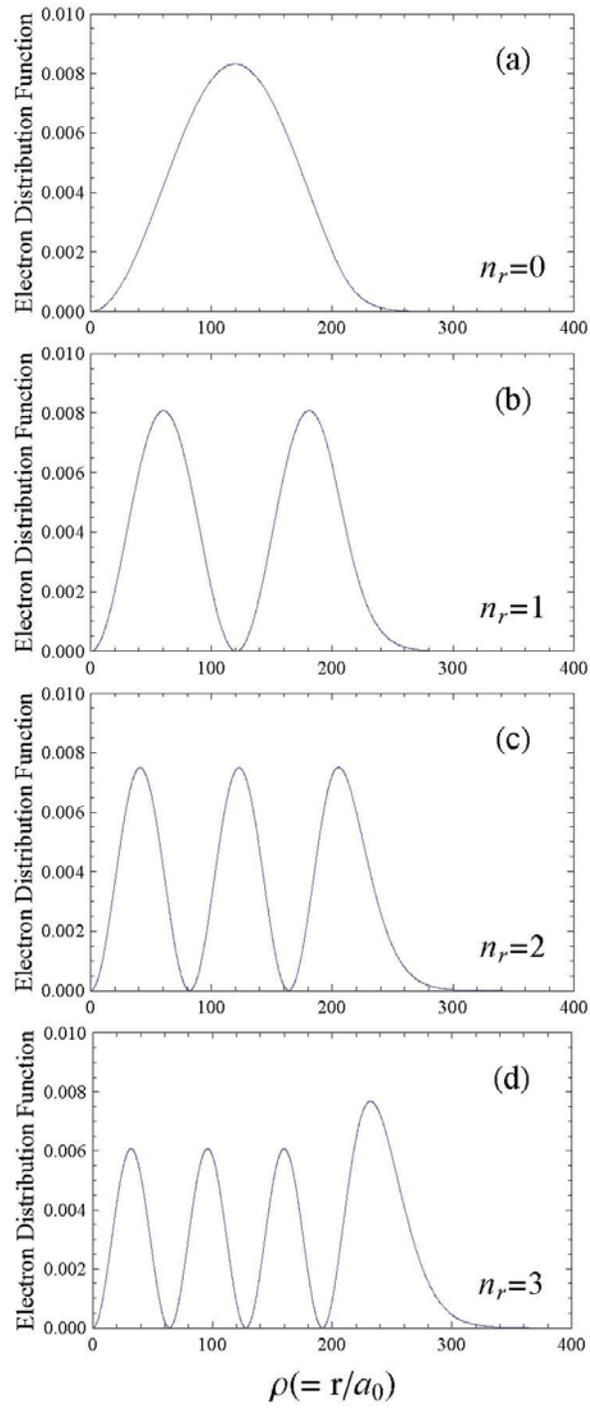

Fig. 1